
\input phyzzx
\catcode`@=11

\newtoks\KUNS
\newtoks\HETH
\newtoks\monthyear
\Pubnum={KUNS~\the\KUNS\cr HE(TH)~\the\HETH}
\monthyear={\monthname,\ \number\year \cr hep-th/9404033}
\KUNS={0000}
\HETH={00/00}
\def\p@bblock{\begingroup \tabskip=\hsize minus \hsize
   \baselineskip=1.5\ht\strutbox \topspace-2\baselineskip
   \halign to\hsize{\strut ##\hfil\tabskip=0pt\crcr
   \the\Pubnum\cr \the\monthyear\cr }\endgroup}
\def\bftitlestyle#1{\par\begingroup \titleparagraphs
     \iftwelv@\fourteenpoint\else\twelvepoint\fi
   \noindent {\bf #1}\par\endgroup }
\def\title#1{\vskip\frontpageskip \bftitlestyle{#1} \vskip\headskip }

\catcode`@=12
\KUNS={1257}
\HETH={94/05}
\monthyear={April, 1994}
\def\1#1{{1 \over {#1}}}
\def\2#1{{{#1} \over 2}}
\def\3#1{{{#1} \over 3}}
\def\4#1{{{#1} \over 4}}
\def\5#1{{{#1} \over 5}}

\def\sss{\scriptscriptstyle}
\def\MP{M_{{\rm P}}}
\def\MGUT{M_{{\rm {\sss GUT}}}}

\def\Tr{{\rm Tr}}
\def\widebar#1{\vbox{\ialign{##\crcr
          \hskip 1.0pt\hrulefill\hskip 1.0pt
          \crcr\noalign{\kern-1pt\vskip0.07cm\nointerlineskip}
          $\hfil\displaystyle{#1}\hfil$\crcr}}}

\def\wt{\widetilde}
\def\NP#1{{\it Nucl}.~{\it Phys}. {{\bf #1}},}
\def\PL#1{{\it Phys}.~{\it Lett}. {{\bf #1}},}
\def\PR#1{{\it Phys}.~{\it Rev}. {{\bf #1}},}
\def\PRL#1{{\it Phys}.~{\it Rev}.~{\it Lett}. {{\bf #1}},}
\def\PTP#1{{\it Prog}.~{\it Theor}.~{\it Phys}. {{\bf #1}},}
\def\RMP#1{{\it Rev}.~{\it Mod}.~{\it Phys}. {{\bf #1}},}
\REF\WIT{
	E.~Witten,
	\NP{B188} 513 (1981); \NP{B202} 253 (1982).
}
\REF\WEI{
	For general review, see;
	S.~Weinberg,
	\RMP{61} 1 (1989).
}
\REF\TSGC{
	M.~Cveti\v c, A.~Font, L.~E.~Ib\'a\~nez,
	D.~L\"ust and F.~Quevedo,
	\NP{B361} 194 (1991),
	and references therein.
}
\REF\GC{
	H.~P.~Nilles,
	\PL{115B} 193 (1983);
\nextline
	S.~Ferrara, L.~Girardello and H.~P.~Nilles,
	\PL{125B} 457 (1983);
\nextline
	M.~Dine, R.~Rohm, N.~Seiberg and E.~Witten,
	\PL{156B} 55 (1985).
}
\REF\TSi{
	K.~Kikkawa and M.~Yamasaki,
	\PL{149B} 357 (1984);
\nextline
	N.~Sakai and I.~Senda,
	\PTP{75} 692 (1986).
}
\REF\DKL{
	V.~Kaplunovsky,
	\NP{B307} 145 (1988);
\nextline
	L.~J.~Dixon, V.~S.~Kaplunovsky and J.~Louis,
	\NP{B329} 27 (1990); \NP{B355} 649 (1991).
}
\REF\IL{
	L.~E.~Ib\'a\~nez and D.~L\"ust,
	\NP{B382} 305 (1992).
}
\REF\BIM{
	A.~Brignole, L.~E.~Ib\'a\~nez and C.~Mu\~noz,
	FTUAM-26/93 (1993).
}
\REF\FIi{
	P.~Fayet,
	\NP{B90} 104 (1975);
\nextline
	P.~Fayet and J.~Iliopoulos,
	\PL{51B} 461 (1974).
}
\REF\FIii{
	W.~Fischler, H.~P.~Nilles, J.~Polchinski, S.~Raby
	and L.~Susskind,
	\PRL{47} 757 (1981).
}
\REF\EN{
	J.~Ellis and D.~V.~Nanopoulos,
	\PL{110B} 44 (1982).
}
\REF\Di{
	M.~Drees,
	\PL{181B} 279 (1986);
\nextline
	J.~S.~Hagelin and S.~Kelley,
	\NP{B342} 95 (1990);
\nextline
	A.~E.~Faraggi, J.~S.~Hagelin, S.~Kelley and D.~V.~Nanopoulos,
	\PR{D45} 3272 (1992);
\nextline
	Y.~Kawamura, H.~Murayama and M.~Yamaguchi,
	DPSU-9303/TU-439 (1993).
}
\REF\KMY{
	Y.~Kawamura, H.~Murayama and M.~Yamaguchi,
	in preparation.
}
\REF\GS{
	M.~B.~Green and J.~H.~Schwarz,
	\PL{149B} 117 (1984).
}
\REF\LSi{
	M.~Dine, N.~Seiberg and E.~Witten,
	\NP{B289} 589 (1987);
\nextline
	W.~Lerche, B.~E.~W.~Nilsson and A.~N.~Schellekens,
	{\it ibid},
	609 (1987).
}
\REF\LSii{
	J.-P.~Derendinger, S.~Ferrara, C.~Kounnas and F.~Zwirner,
	\NP{B372} 145 (1992);
\nextline
	G.~Lopes~Cardoso and B.~Ovrut,
	\NP{B369} 351 (1992);
\nextline
	\NP{B392} 315 (1993);
\nextline
	P.~Mayr and S.~Stieberger,
	\NP{B412} 502 (1994).
}
\REF\EFF{
	Y.~Kazama and Y.-P.~Yao,
	\PR{D25} 51 (1982);
\nextline
	L.~Hall, J.~Lykken and S.~Weinberg,
	\PR{D27} 2359 (1983).
}
\REF\IB{
	For related issues, see;
	L.~E.~Ib\'a\~nez,
	\PL{B303} 55 (1993).
}
\REF\RBS{
	K.~Inoue, A.~Kakuto, H.~Komatsu and S.~Takeshita,
	\PTP{68} 927 (1982);
\nextline
	L.~E.~Ib\'a\~nez and G.~G.~Ross,
	\PL{110B} 215 (1982).
}
\REF\SU{
	P.~Ginsparg,
	\PL{197B} 139 (1987);
\nextline
	T.~Banks, L.~J.~Dixon, D.~Friedan and E.~Martinec,
	\NP{B299} 613 (1988).
}
\REF\CYii{
	E.~Witten,
	\NP{B268} 79 (1986);
\nextline
	J.~Distler and B.~Greene,
	\NP{B304} 1 (1988).
}
\REF\DUAL{
	S.~Ferrara, L.~Girardello, T.~Kugo and A.~Van~Proeyen,
	\NP{B223} 191 (1983).
}
\REF\FIiii{
	J.~J.~Atick, L.~J.~Dixon and A.~Sen,
	\NP{B292} 109 (1987).
}
\REF\PCHH{
	J.~Polonyi, Budapest preprint KFKI-93 (1977);
\nextline
	M.~Claudson, L.~J.~Hall and I.~Hinchliffe,
	\PL{130B} 260 (1983).
}
\REF\OB{
	A.~Font, L.~E.~Ib\'a\~nez, F.~Quevedo and A.~Sierra,
	\NP{B331} 421 (1990).
}
\REF\FLIP{
	I.~Antoniadis, J.~Ellis, J.~S.~Hagelin and D.~V.~Nanopoulos,
	\PL{B205} 459 (1988); \PL{B231} 65 (1989).
}
\REF\ECi{
	G.~D.~Coughlan, W.~Fischler, E.~W.~Kolb, S.~Raby
	and G.~G.~Ross,
	\PL{131B} 59 (1983);
\nextline
	T.~Banks, D.~Kaplan and A.~E.~Nelson,
	\PR{D49} 779 (1994).
}
\REF\EXTRA{
The simplest way would be to consider models in which
the VEV of $\chi$ triggers the extra $U(1)$ breakings.
Then one should include the extra $\Lambda_D$'s.
}
\REF\ECii{
	E.~Cohen, L.~Ellis, K.~Enqvist and D.~V.~Nanopoulos,
	\PL{161B} 85 (1985);
\nextline
	J.~Ellis, D.~V.~Nanopoulos and M.~Quir\'os,
	\PL{174B} 176 (1986);
\nextline
	B.~de~Carlos, J.~A.~Casas, F.~Quevedo and E.~Roulet,
	\PL{B318} 447 (1993).
}

\titlepage

\title{
	Scalar Mass and Cosmological Constant
	induced by ``Anomalous" $U(1)$ $D$-term
}
%

\author{
	Hiroaki~NAKANO
\foot{
	Present address:
	Department of Physics,
	Niigata University,
	Niigata 950-21, Japan
} }

\address{
	Department of Physics,~Kyoto University
	\break
	Kyoto~606,~JAPAN
}

\abstract{
When the supersymmetric theory contains
the ``anomalous" $U(1)$ gauge symmetry
with Green-Schwarz anomaly cancellation mechanism in 4 dimensions,
its Fayet-Iliopoulos $D$-term generates
non-universal scalar masses and the positive cosmological constant
after the supersymmetry breaking.
Both give the new contributions to the known results from $F$-term.
Our mechanism is naturally realized in many string models
and in some cases, leads to remarkable cancellations
between $F$- and $D$-term contributions, providing
the universal scalar mass and vanishing cosmological constant.
We illustrate how such a possibility can arise
by taking a simple orbifold example.
}

\endpage
\sequentialequations

Supersymmetry (SUSY) provides us with an attractive picture for
the physics above the weak scale $m_Z$ up to the Planck scale $\MP$.
Despite the beautiful structure,
we have no convincing evidence yet that
such an idea is realized in Nature.
This is primarily due to our lack of satisfactory mechanism
for dynamical SUSY breaking\rlap.\refmark{\WIT}
It should explain the origin of the mass hierarchy $m_Z \ll \MP$
and the reason why the cosmological constant $\Lambda$
vanishes\refmark{\WEI} after the SUSY breaking.
It should also lead to the predictive low-energy theory
to be experimentally checked;
without it, we are faced with
general set of soft SUSY-breaking parameters.

Here we have little to say about the SUSY-breaking mechanism.
We assume that
some mechanism primordially generates
the observable SUSY-breaking scale $m_{3/2} \sim m_Z$.
In fact, there has been recent development in this area;
in string theory, for instance,
it was argued\refmark{\TSGC} that
the hidden-sector gaugino condensation\rlap,\refmark{\GC}
combined with the stringy quantum symmetry,
target-space duality\rlap,\refmark{\TSi}
generates the nonperturbative superpotential,
which may break the SUSY in $F$-term sector.
Soft breaking parameters were also
calculated\rlap.\refmark{\DKL,\IL,\BIM}

So far, much effort has concentrated only on $F$-term sector.
After early attempts\rlap,\refmark{\FIi,\FIii}
little attention has been paid to $D$-term sector.
Our purpose here is to address its possible role
in the phenomenological issue related to scalar masses
and also in the cosmological constant problem.
We find that if the theory contains
an ``anomalous'' $U(1)$ gauge symmetry,
its Fayet-Iliopoulos (FI) $D$-term leads to interesting,
remarkable consequences once the SUSY breaking
in $F$-term sector is incorporated.

Low-energy phenomenology requires\refmark{\EN} to some extent
the ``universality" of soft-breaking scalar masses;
motivated by ``minimal" supergravity,
most of phenomenological analyses assume that
the scalar masses are all equal at $\MP$,
and deal only with the radiative corrections which
invalidate the equality at a scale lower than $\MP$.
However, being not the logical consequence of local SUSY,
such an assumption may well be violated.
Actually the universality does not holds
in many string-model calculations\rlap.\refmark{\IL}
Moreover some authors point out\refmark{\Di,\KMY}
in the context of grand unification
that even if it holds primordially at $\MP$,
when an $U(1)$ is broken at $\MGUT$,
the universality is lost due to its $D$-term,
provided that
SUSY-breaking effects are properly incorporated
and that the primordial universality is already violated
at $\MGUT$ by radiative corrections above $\MGUT$.
The appearance of such $D$-term mass splitting, however, seems to
depend on the details of the superpotential above $\MGUT$
which gives rise to the requisite radiative corrections.
The last condition was necessary essentially because
the $U(1)$ generator, embedded in a simple group, is traceless.
The situation drastically changes
when the theory contains the so-called ``anomalous" $U(1)$;
an $U(1)$ gauge symmetry with nonvanishing anomaly,
when calculated from massless spectrum alone,
which is actually canceled
via 4d Green-Schwarz (GS) mechanism\rlap.\refmark{\GS}
Then the $U(1)$ generator $Q$ is no longer traceless $\Tr Q\not=0$
and the $D$-term splitting arises independently of superpotential.

Let us first adopt a simple model to illustrates the idea.
Apart from the unspecified sector
responsible for the SUSY breaking,
the model contains an anomalous-$U(1)$ vector multiplet,
a ``dilaton-axion" (linear)multiplet $L$
responsible for GS mechanism\rlap,\refmark{\LSi,\LSii}
and set of chiral multiplets whose scalar components we denote as
${\Phi}_I=(\phi_0 \equiv \chi,{\phi}_i)$
with charge $Q_I=(Q_0, Q_i)$.
We assume that only $Q_0$ takes the opposite sign to $\Tr Q$,
and for simplicity, assume canonical kinetic terms
and vanishing superpotential for chiral fields.
We start with the situation with the unbroken SUSY
and vanishing FI $D$-term at tree level.
Its generation at one-loop\refmark{\FIii,\LSi}
leads to the {\it non-quartic} scalar potential
$$
V^{(0)}({\Phi})
 = \2{g^2} {\Big[} M^2\Tr Q + \sum_I Q_I|{\Phi}_I|^2 {\Big]}^2
\eqn\eqPOTi$$
where $g^2$ is a gauge coupling constant and
$M$ is a cutoff for quadratic divergence in FI term.
Under our assumptions,
this potential shows the global minimum $\chi={\chi}^{(0)}$ at
$$
M^2\Tr Q + Q_0|{\chi}^{(0)}|^2 = 0 \ .
\eqn\eqVEVi$$
Nothing happens at this stage;
the SUSY remains unbroken,
and the vector multiplet becomes massive
and decouples\rlap.\refmark{\LSi}

We now assume\refmark{\KMY} that the SUSY breaking
results in the soft breaking masses $m_I^2(>0)$
of the scalar ${\Phi}_I$, of order $m_{3/2}^2 \ll M^2$.
The $m_I^2$ may or may not be universal.
Then the scalar potential takes the form
$$
V(\Phi) = V^{(0)}(\Phi) + \sum_Im_I^2|{\Phi}_I|^2 \ .
\eqn\eqPOTii$$
We integrate out the heavy field $\chi$ to obtain
the effective potential\refmark{\EFF}
of the light fields ${\phi}_i$.
A minimization with respect to $\chi$ gives
$$
M^2\Tr Q + Q_0|\chi|^2
 = - { m_0^2 \over Q_0g^2} - \sum_i Q_i|{\phi}_i|^2
\eqn\eqVEVii$$
which we solve as $\chi=\chi({\phi}_i)$.
Observe the order $m_{3/2}^2$ shift\refmark{\KMY,\EFF}
from $|{\chi}^{(0)}|^2$
in the vacuum expectation value (VEV) of $\chi$.
The effective potential
$V_{{\rm eff}}(\phi)= V{\big(}\chi({\phi}_i),{\phi}_i{\big)}$ is
$$
V_{{\rm eff}}(\phi)
 =   {\Tr Q \over -Q_0}m_0^2M^2
   - { m_0^4 \over 2Q_0^2g^2}
   + \sum_i {\Big[}m_i^2
   + {Q_i \over -Q_0} m_0^2{\Big]}|{\phi}_i|^2
\eqn\eqPOTiii$$
We thus find the $D$-term contributions to scalar mass
${\big(}m_i^2{\big)}_D = Q_im_0^2/(-Q_0)$
which are generally {\it non-universal},
and also the {\it positive} cosmological constant
$\Lambda_D$ of order $m^2_0M^2$.
${\big(}m_i^2{\big)}_D$ does not depend on a cutoff $M^2$
while the $\Lambda_D$ does.

What are phenomenological implications of $D$-term mass splitting?
$\Tr Q\not=0$ implies that we may assign $U(1)$ charge $Q_i$ at will,
and the primordial universality will, if any, be completely broken.
This may lead to a disaster,
or to an interesting scenario;\refmark{\IB}
it might split the masses of Higgses
in SUSY standard model,
breaking the electroweak symmetry around $m_{3/2}$.
This can be an alternative to
widely-studied radiative breaking
scenario\rlap.\refmark{\RBS}
Anyway, a moral of this simplest model seems to be that
we can have almost any type of scalar masses as we desire
(unless gaugino masses dominate them over).

The situation again changes
if we suppose that the theory be embedded into the supergravity
in which we should take care of other mixed-type anomalies.
It is likely that the absence of
K\"ahler, $\sigma$-model as well as superconformal
anomalies\refmark{\LSii} forces the theory
to be embedded into a string model where
the world-sheet modular invariance will
guarantee the full consistency.
Such an embedding will be the unique way
to make sense of an $U(1)$
without an unification group\rlap.\refmark{\SU}
Hereafter we focus on (0,2)-string models\refmark{\CYii}
with an anomalous $U(1)$.

Let us briefly recall the relevant things
about generic string models\rlap.\refmark{\TSGC,\IL,\BIM}
First,
the target-space duality symmetry $\Gamma$ constrains
the possible nonperturbative superpotential $W(S,T)$
of the dilaton multiplet $S$ (dual\refmark{\DUAL} to $L$)
and overall K\"ahler moduli multiplet $T$.
We assume the diagonal K\"ahler potential $K$ for simplicity.
The resulting scalar potential
generically yield\refmark{\TSGC}
the {\it negative} cosmological constant
$$
\Lambda_F = m_{3/2}^2\MP^2
	{\Big[} | \, {\wt F}_S|^2 + |{\wt F}_T|^2 - 3 \, {\Big]}
	{\Big(}<0{\Big)} \ .
\eqn\eqCCF$$
${\wt F}_{S,T}$ are essentially auxiliary fields of $S$ and $T$,
whose nonzero VEV's break the SUSY
and are parametrized\refmark{\BIM}
by the $\Lambda_F$, gravitino mass $m_{3/2}^2=e^K|W|^2$ and
Goldstino angle, $\tan \theta =|{\wt F}_S/{\wt F}_T|$.
We treat them as free parameters
since the problem of dynamical SUSY-breaking is not settled.
Next,
massless matter fields $\Phi_I$ are classified\refmark{\DKL,\IL}
according to an integer $n_I$ called ``modular weight",
which characterizes the transformation property under $\Gamma$.
In simple orbifold cases,
it takes $n=-2~(-1)$ for (un)twisted matter.
It was argued\refmark{\IL} that the scalar masses are not universal,
but depend on the modular weight as
${\big(}m_I^2{\big)}_F = {\wt m}_0^2 + n_I{\wt m}^2_1$,
which depend on three parameters $\Lambda_F$, $m_{3/2}^2$, $\theta$
($d=0$ at tree level) through\refmark{\BIM}
$$
{\wt m}_0^2 = m_{3/2}^2 + { \Lambda_F \over \MP^2} \ , \qquad
{\wt m}_1^2 = {\Big[}m_{3/2}^2 + { \Lambda_F \over 3\MP^2}{\Big]}
	{\cos^2 \theta \over 1-d}\ {\Big(}>0{\Big)} \ .
\eqn\eqMASSFi$$

We now add the $D$-term contributions.
We neglect the possible threshold corrections\refmark{\DKL}
to the $D$-term \eqPOTi.
As for the cosmological constant,
keeping only the dominant term in eq.~\eqPOTiii,
$$
c\,\Lambda_D
	= m_{3/2}^2\MP^2{\big(}1+n_0\cos^2\theta{\big)}
	+ \Lambda_F{\big(}1+\3{n_0}\cos^2 \theta{\big)}
\eqn\eqCCD$$
where $c \equiv (-Q_0/\Tr Q) \MP^2/M^2$.
[$M^2=\MP^2 / (192\pi^2)$
in string\rlap.\refmark{\FIiii}]
This is positive for $m_0^2={\wt m}_0^2+n_0{\wt m}_1^2>0$.
In a given model with $\Lambda_F<0$,
we can always find a value of $m_{3/2}^2$ so that
{\it the total cosmological constant vanishes} $\Lambda_{F+D}=0$.
This process may well involve fine-tuning, but
{\it it can be done without any other $F$-term contributions
nor complicated K\"ahler potential}\rlap.\refmark{\PCHH}
A similar $\Lambda_D$ may arise,
depending on the matter superpotential,
even if the $U(1)$ is not anomalous.
With an anomalous $U(1)$,
it always arises and as we will see,
is intimately related to the scalar-mass universality.

As for the scalar masses,
we take a model of ref.~[\OB]
with a left-right symmetric gauge group
as an interesting example.
[This model was obtained from $Z_3$-orbifold with
suitable gauge embedding and two Wilson-lines.]
An anomalous $U(1)$ comes from
the second $E_8$ of the heterotic string.
Before the anomalous-$U(1)$ breaking,
$\Tr Q=+864$ and a single type of fields $\chi$
($u$ in ref.~[\OB]) with $Q_0=-12$ and $n_0=-1$
develops the VEV $\chi^{(0)}$ to cancel the $D$-term.
After the breaking,
surviving fields $\phi_i$ show an interesting feature;
the $U(1)$ charge $Q_i$ is ``proportional" to
the modular weight $n_i$.
In visible sector,
untwisted matters (other than $\chi$) are neutral,
and massless twisted matters have charge $Q_i= + 4$.
This will enable us to have the restored universality.
Taking more general case in mind, let us make an ansatz
$$
{Q_i \over Q_0} = {n_i-b \over a} \ .
\eqn\eqANSATZ$$
Combining $F$- and $D$-contributions,
$$
{\big(}m_i^2{\big)}_{F+D}
 = {\Big[} {\wt m}_0^2 + b{\wt m}_1^2 {\Big]}
 + {b-n_i \over a}{\Big[}{\wt m}_0^2
		+ {\big(}n_0-a{\big)}{\wt m}_1^2{\Big]} \ .
\eqn\eqMASSFD$$
If we tune the Goldstino angle $\theta$ so that
$m_0^2 - a{\wt m}_1^2
= {\wt m}_0^2 + {\big(}n_0 - a{\big)}{\wt m}_1^2 = 0$,
then the $Q_i$-dependence in ${\big(}m_i^2{\big)}_D$ cancels out
the $n_i$-dependence in ${\big(}m_i^2{\big)}_F$
and we have the universal scalar mass
$m^2_{F+D}= {\big(} a+b-n_0 {\big)} {\wt m}_1^2$.
[Having $a>0$
will be enough to have milder dependence on modular-weights
even {\it without} the ansatz \eqANSATZ.]
Finally we present a simultaneous solution
to the vanishing $\Lambda_{F+D}$ and universal $m^2_{F+D}$
for a given $\Lambda_F<0$.
It is ($d=0$ at tree level)
$$
{m_{3/2}^2\MP^2 \over -\Lambda_F}
= 1 + {\Big(}1-{n_0 \over a}{\Big)}c \ , \qquad
{1 - d \over \cos^2\theta}
= {2a \over 3c} + {\Big(}1-{n_0 \over a}{\Big)}a
\eqn\eqSOLi$$
with the universal scalar mass
$$
m^2_{F+D} = { (a+b-n_0)c \over a + (a-n_0)c } m_{3/2}^2 \ .
\eqn\eqSOLii$$
In our example ($a=3$, $b=n_0=-1$, $c=8\pi^2/3$),
$m_{3/2}^2 \sim 2(4\pi /3)^2(-\Lambda_F)/\MP^2$,
$\cos^2 \theta \sim 1/4$ and $m^2_{F+D} \sim m^2_{3/2}/4$
(in the limit $c \rightarrow \infty$).

We have found that after the SUSY breaking,
the FI $D$-term of anomalous-$U(1)$ generates
new contributions to the scalar masses and cosmological constant
which have not been noticed previously.
The former generally breaks the universality of scalar masses
independently of the superpotential and may give an impact
on phenomenology of low-energy SUSY.
On the other hands in some string models,
it restore at least partially the broken universality.
Also the latter provides a chance
to automatically cancel the cosmological constant.
It is worth examining other string models
(such as flipped $SU(5) \times U(1)$\refmark{\FLIP})
to clarify the generality of our mechanism.
Also the complete analysis including other soft-breaking parameters
as well as one-loop corrections\refmark{\DKL,\LSii} are desired.
Finally
as in the generic supergravity models\rlap,\refmark{\ECi,\ECii}
our model may suffer from the ``entropy crisis",
the cosmological problem related to the ``axions".
Its solution\refmark{\EXTRA} might require
the better understanding for the mixing of
the Goldstone mode (Im$\chi$) and axions (Im$S$, Im$T$),
or require\refmark{\ECii} the rather large $m_{3/2}=O(10)$ TeV,
in which case the phenomenological requirement of
the scalar-mass universality disappear as well.
Works along these lines are in progress.

\vskip 0.5cm

The author would like to thank
K.~Inoue, Y.~Kawamura, T.~Kobayashi and M.~Yamaguchi
for invaluable suggestions,
and T.~Kugo and S.~Yahikozawa
for stimulating discussions.

\refout

\bye